\documentclass{iaus}
\usepackage{graphicx}
\usepackage{amsmath}

\title[Search for extrasolar planets with high-precision relative astrometry] 
{Search for extrasolar planets with high-precision relative astrometry by ground-based and single-aperture observations}

\author[T. Roell \& A. Seifahrt \& R. Neuh\"auser]   
{Tristan Roell$^1$
 \and Andreas Seifahrt$^{1,\, 2}$ \and Ralph Neuh\"auser$^1$}

\affiliation{
$^1$Astrophysikalisches Institut und Universit\"ats-Sternwarte Jena, \\ email: {\tt troell@astro.uni-jena.de}\\ email: {\tt rne@astro.uni-jena.de} \\[\affilskip]
$^2$Institut f\"ur Astrophysik, G\"ottingen\\ email: {\tt seifahrt@astro.physik.uni-goettingen.de}
}

\pubyear{2008}
\volume{249}  
\pagerange{119--126}
\jname{Extrasolar planets}
\editors{A.C. Editor, B.D. Editor \& C.E. Editor, eds.}
\begin{document}

\maketitle

\begin{abstract}
We present our search program for substellar companions using high-precision relative astronomy. Due to its orbital motion around the star, an unseen substellar companion would produce a periodic "wobble" of the host star, which is the astrometric signal of the unseen companion. By measuring the separation between the components of stellar double and triple systems, we want to measure this astrometric signal of a possible unseen companion indirectly as a relative and periodic change of these separations. Using a new observation mode (the "cube-mode") where the frames were directly saved in cubes with nearly no loss of time during the readout, an adaptive optics system to correct for atmospheric noise and an infrared narrow band filter in the near infrared to suppress differential chromatic refraction (DCR) effects we achive for our first target (the double star HD 19994) a relative precision for the separation measurements of about $100\ldots 150\,\mu as$ per epoch. To reach a precision in the $\mu as$-regime, we use a statistical approach. We take several thousand frames per target and epoche and after a verification of a Gaussian distribution the measurement precision can be calculated as the standard deviation of our measurements divided by the square root of the number of Gaussian distributed measurements. Our first observed target is the stellar binary HD 19994 A \& B, where the A component has a known radial velocity planet candidate.

\end{abstract}

\firstsection 

\section{Introduction}

Up to now, most of the extrasolar planets have been detected with the radial velocity (RV) technique. But due to the unknown inclination angle $i$ this technique just yields the lower mass limit $M\sin{i}$ and not the true mass of the substellar companion. Therefore, all radial velocity planets should be regarded as planet candidates, until their true mass is determined. In contrast to the radial velocity technique, astrometry yields the inclination angle by measuring the astrometric signal of the substellar companion and hence its true mass. Until now, three radial velocity planet candidates could be confirmed with absolute astrometry using the Fine Guiding Sensor (FGS) of the Hubble Space Telescope (HST), GJ 876 b by \cite{benedict_02_gj876}, 55 Cancri d by \cite{arthur_04_55cnc} and Epsilon Eridani b by \cite{benedict_06_eridani}. 
Recently, \cite{bean_07} measured the astrometric signal of the radial velocity planet candidate HD 33636 b ($M\sin{i}=9.3\,M_J$) and obtained a value for the true mass of the companion of $M=142\pm 11\,M_J$, thus it is a low mass star. This clarifies the importance of astrometric follow up observations to determine the true mass of radial velocity planet candidates. 
The mass is one of the most important stellar and substellar parameters and plays a key role in our understanding of the distribution, forming and evolution of substellar objects. Besides all other methods to determine the mass, which are using theoretical predictions (like evolutionary models), astrometry is a method, which is independent from theoretical assumptions (hence also from theoretical uncertanties) by measuring the dynamical mass of the objects.

Besides the confirmation of RV planet candidates, we also want to search for unknown substellar companions in stellar double and triple systems. This kind of search program will provide new insights into the formation and evolution of extrasolar planets in multiple systems, which is important due to the fact that more than $50\%$ of all stars are members of a multiple system.

\section{Observation method}

To reach a precision comparable to the HST observation, we observe double and triple stars and measure the separation between all stellar components, thus using relative astrometry. In the case of an unseen substellar companion, we would measure the astrometric signal indirectly as a relative and periodic change in the separations.
The quest of measuring the astrometric signal of a substellar companion needs a carfully handling of all noise sources, which are among others atmospheric noise, photon noise, background noise, readout-noise and DCR.
Our observations on the southern hemisphere are done with the 8.2 meter telescope UT4 of the ESO Very Large Telescope (VLT) and the NACO S13 (NAOS-CONICA) infrared camera.
Using the adaptive optics system NAOS (Nasmyth Adaptive Optics System) we correct for atmospheric turbulences and by using a narrow band filter centered in the near infrared ($\lambda_{cen}=2.17\,\mu m$) we suppress DCR effects.
Due to the use of the double-correlated readout mode, we suppress readout noise and by choosing a suitable exposure time (to reach a high signal to noise ratio) we can neglect photon and background noise. 
The pixel scale of the used detector (NACO S13) is about $13.25\, mas$, which means a Field of View (FoV) of about $14''\times 14''$. The guide star for the AO system is always one of our stellar components. The separation of our observed multiple systems is typically four arcseconds. Hence, the angular separation is (with normal seeing conditions) always smaller than the isoplanatic angle. 

Furthermore (besides the use of relative astrometry), we use a new observation mode, called "cube-mode". This mode saves frames directly in a cube and thus has nearly no loss of time during the readout. With the minimal exposure time of $0.35$ seconds using the double-correlated readout mode it is possible to obtain 2500 frames within 15 minutes. 

\begin{figure}
	\begin{center}
		\includegraphics[width=13cm]{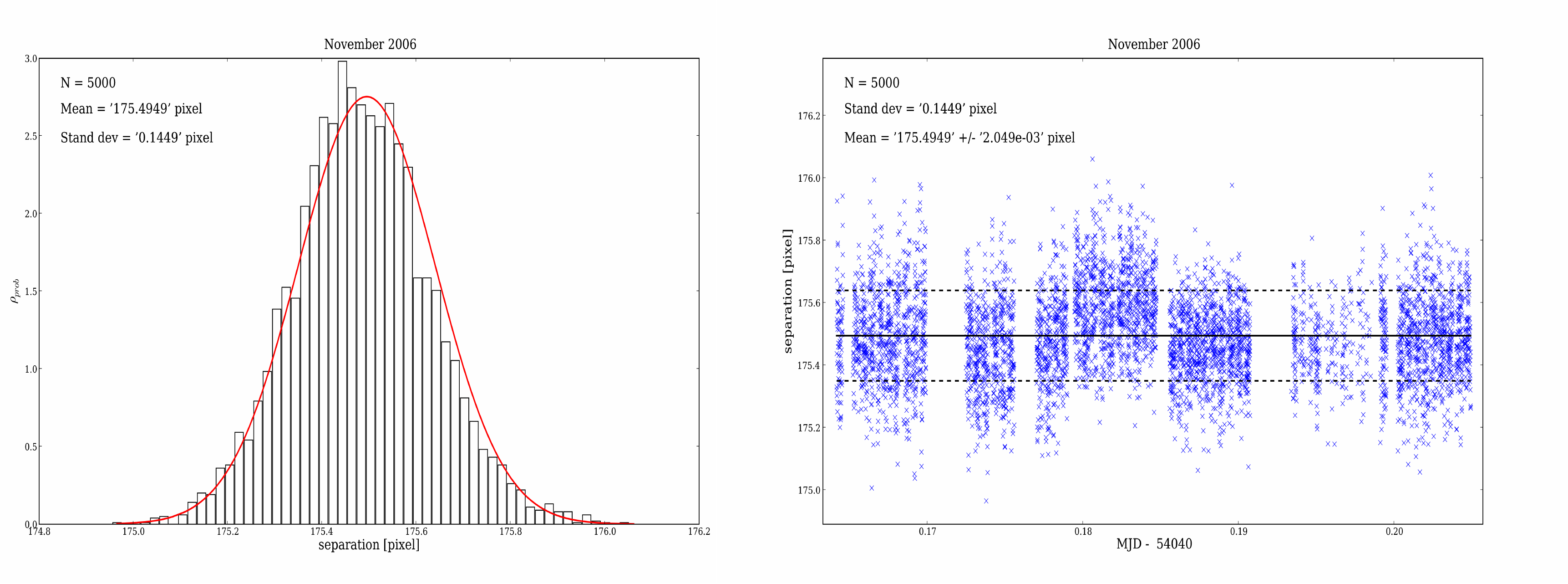}
		\includegraphics[width=13cm]{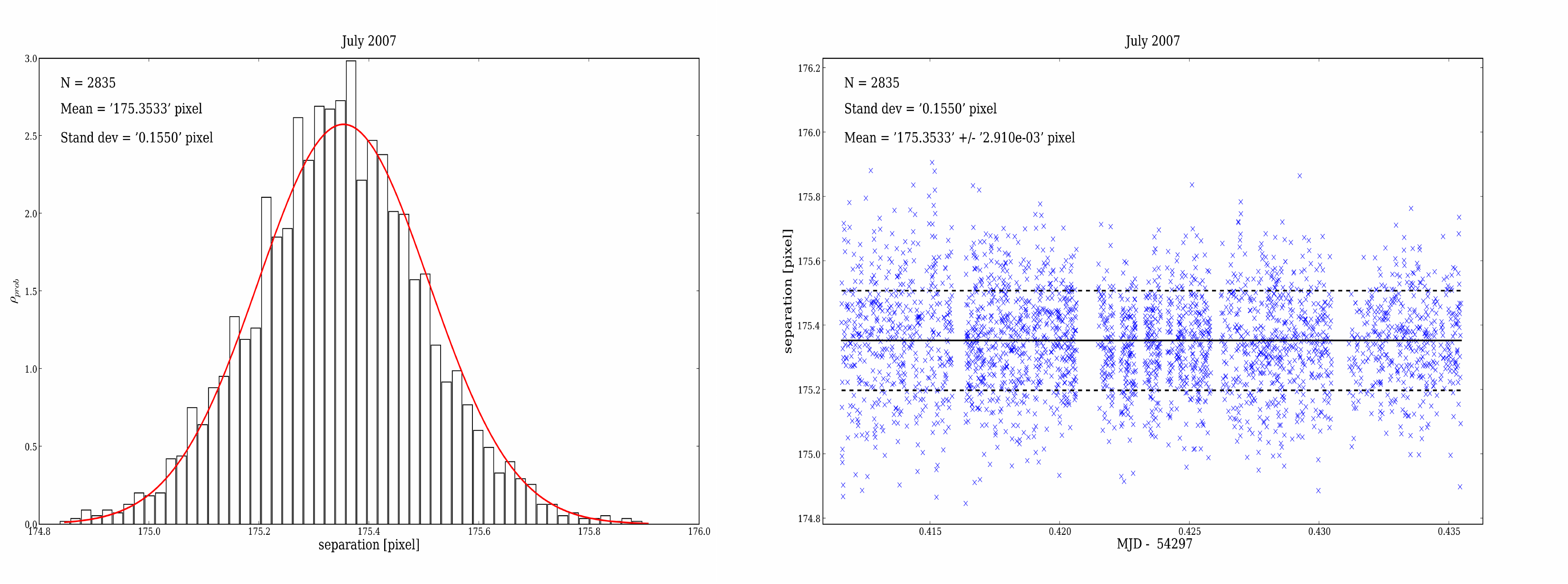}		
		\caption{Separation measurements (right) and the Gaussian distribution of the measurements (left) of the stellar binary HD 19994 from 2006 (top) and 2007 (bottom)}
		\label{fig_1}
	\end{center}
\end{figure}

The following statistical principle is similiar to the method of measuring the radial velocity with hundreds of spectral lines to reach a higher precision, which is used in the radial veocity technique. In our astrometric case we measure the separation between all stellar components in every frame and obtain several thousand measurements of the same separation. After a verification of Gaussian distributed measurements (done with a Kolmogorov-Smirnov-Test) and a two sigma clipping (to eject frames with bad quality, due to the non-constant performance of the AO system and the dynamical seeing behaviour), the measurement precision ($\Delta_{meas}$) can be calculate as the standard deviation of the measurements ($\sigma_{meas}$) divided by the square root of the number of Gaussian distributed measurements ($N$), $\Delta_{meas}=\dfrac{\sigma_{meas}}{\sqrt{N}}$ (see Fig. \ref{fig_1}).

But we have to keep in mind that the above value is just the measurement precision and includes only the statistical distributed error sources. To determine the systematic error sources, which affect in the case of relative astrometry the pixel scale and the position angle, we need a calibration system (see next section).

\section{Calibration}

Because we are dealing with relative astrometry we do not need an absolute astrometric calibration of our data, but we have to monitor the stability of our pixel scale to correct our measurements for possible variations of the pixel scale. 
The "normal way" of relative astrometric calibration is to use a Hipparcos binary system. This method results in the case of the NACO S13 camera in a pixel scale of typically $13.25 \pm 0.05\,mas$ per pixel (see \cite{neuh_05}). We do not need the absolute value of the pixel scale, but we have to determine changes very precisly. Hence, what we need is a calibration system with a high and known intrinsic stability (higher than a Hipparcos binary). We choose the old globular cluster 47 Tuc as our calibration cluster for targets on the southern hemisphere. The reasons are a high number of "calibration stars" in the FoV of the S13 NACO camera and a known intrinsic stability. \cite{laughlin_06} determined the transversal velocity dispersion of the 47 Tuc cluster members and obtained a value of about $630\,mas$ per year. To monitor the pixel scale we take hundreds of frames of 47 Tuc per epoch, measure the separation from each star to each star and compute the mean of all these separation measurements on every single frame. This mean of the separations represents the relative alignment of all observed cluster members and should have, within the errors (intrinsic instability and measurement errors), the same value in every epoch. Using a Monte-Carlo-Simulation with our observed cluster members and a Gaussian distributed transversal velocity dispersion of $630\,mas$ per year, we obtain an intrinsic stability of our used calibration cluster of about $3/100000$ per pixel and year, which results for a given pixel scale of $13.25\,mas$ for the S13 camera an intrinsic stability of about $0.4\,\mu as$ per pixel and year. This means, with 47 Tuc we are able to determine a change in the pixel scale down to $0.5\ldots 1\,\mu as$ per pixel and year (including the typical measurement errors of the 47 Tuc cluster members). Due to the fact, that we are using a fixed and given ``reference pixel scale" of $13.25\,mas$ in our first epoch, we can not determine the absolute value of the pixel scale, but we are able to detect changes between the epoches very precisely. 

Furthermore, we have to consider that the observed and measured change in the separation is a product of the orbital movement of the unseen companion and the stellar system (double or triple system). Regarding the timescale of both influences (hence the orbital periods of the substellar unseen companion and the seen stellar companion) we see a large difference. The orbital period of stellar binaries with a separation of about three arcseconds is typacilly more then thousend years. Astrometric search programs are most sensitive for the intermediate semi-major-axis regime, which means a period of about one year up to about three years. Thus, the influence of the stellar binary during our search program is nearly linear or just slightly curved. The advantage of a large difference in the orbital periods are that we are not depend on the exact orbital elements of the stellar binary and we are able to distinguish both influences due to the different timescale.
There are two possibilities to determine the influence of the stellar binary. Either to add the stellar influence as an additional free paramater in the final fit or to determine the influence directly by measuring the separation of the stellar binary at two (linear influence) or three (sligthly curved influence) epoches where the substellar companion has the same orbital position. The second possibility can be applied for RV planet candidates (the orbital period of the candidate is well known) but to search for unknown companions, we have to use the first method thus to consider the nearly linear stellar influence in the final fit.

Our first observed stellar binary (HD 19994) has a separation of about $2.5$ arcsec and an orbital period of about 1500 years. \cite{mayor_04} discovered a RV planet candidate around the A component with $M\,\sin{i}=1.68\,M_J$, $a=1.4$ AU and a period of about $535$ days. For HD 19994 we achive a total relative precision of $100\ldots 150\,\mu as$ per epoch. The expected change in the separation (due to the astrometric signal of the planet candidate) depends on the orientation of both orbits (especially from the difference in the inclination angle between the planetary and the double star orbit) and is about $450\,\mu as$ for $\Delta i \approx 50^\circ$ ($M_{true}\approx 5\,M_J$, $2\,\sigma$ detection). In the case of an astrometric non-detection we are able to exclude differences in the inclination angle of more than $\Delta i \approx 65^\circ$ ($3.5\,\sigma$ detection) and thus to exclude true masses of more than $9\,M_J$.

\end{document}